# Innovative NiAl Electrodes for Long Term, Intermediate High Temperature SAW sensing applications using LiNbO$_3$ substrates

Jordan MAUFAY[1,2], Baptiste PAULMIER[2], Mélanie EMO[2], Ulrich YOUBI[2], Esther MBINA[2], Thierry AUBERT[1], Ninel KOKANYAN[1], Sami HAGE-ALI[2], Michel VILASI[2] and Omar ELMAZRIA[2]

[1]Université de Lorraine, CentraleSupélec, LMOPS, F-57000 Metz, France
[2]Université de Lorraine, CNRS, IJL UMR 7198, F-54000 Nancy, France

*Abstract*— Wireless SAW reflective delay line (R-DL) technology is very powerful to carry out remote measurements of various parameters under harsh environments, while enabling the identification of a given sensor among several of them. However, R-DL technology is currently limited to 350°C for long-term applications, likely because of aluminium electrodes oxidation and/or congruent lithium niobate segregation process. In this study, an innovative alloy, namely NiAl, is investigated as an alternative to Al to make R-DLs able to withstand high temperatures up to 500°C on the long term. Indeed, NiAl gathers, in the bulk state, all the necessary properties (fairly low electrical resistivity and density, high melting temperature and resistance to oxidation). The study also examines the extent of the congruent LiNbO$_3$ segregation process, to determine its impact on the NiAl/LiNbO$_3$ R-DLs performances.

The obtained results are very promising. NiAl electrodes self-passivate during the first 50h of annealing at 500°C: 20 nm-thick Al$_2$O$_3$ layers form at the surface and in between the electrodes and the substrate, protecting the remaining NiAl layer from further oxidation. Besides, the segregation process occurs mainly in the same time. It is located in the first 150-200 nm of the substrate. Both phenomena have no significant impact on the performance of NiAl/LiNbO$_3$ R-DLs working at 433 MHz. Continuous in-situ electrical monitoring of such devices shows a standard deviation of the operating frequency of only 1.04 ppm during an annealing process of 250h at 500°C. Moreover, the time-resolved S$_{11}$ response of the device at the end of this treatment is not degraded at all. Thus, 433 MHz NiAl/LiNbO$_3$ R-DL sensors can operate with high fidelity for at least 10 days at 500°C under air atmosphere, and there are strong signs that their lifetime is actually much longer.

*Index Terms*— SAW; High-temperature sensor; Wireless; Batteryless; Lithium niobate; NiAl alloy thin films.

## I. INTRODUCTION

In order to meet the challenge of high-temperature sensing in harsh environments, surface acoustic wave (SAW) sensors are very strong candidates since they are fully passive (i.e. they need neither battery nor embedded electronics to be wirelessly interrogated). In particular, SAW reflective delay lines (R-DLs) are of high interest since R-DL sensors can be unambiguously identified through their coded time-resolved response. In other terms, R-DL technology enables the achievement of SAW sensors networks [1]. Sensor identification through pulse position coding requires both high amplitude and short time pulses, and hence, for the latter, a large bandwidth. Practically, for SAW sensors, only the 915 MHz or 2.45 GHz ISM bands possess an adequate one. These conditions make necessary the use of a piezoelectric substrate with a large electromechanical coupling coefficient (K$^2$) of several percent. Thus, the piezoelectric materials known for their stability at very high temperature, such as langasite [2,3] or aluminium nitride [4], having a low electromechanical coupling (< 0.5%), are unsuitable for the R-DL technology. For these reasons, high K$^2$ LiNbO$_3$ substrates have been the base of the market of SAW wireless identifiable high-temperature sensors for more than a decade [5]. Operating temperatures of 300-350°C are sustainable by these sensors for several days, but the lifetime of the sensors in [6] was found to decrease very quickly with the temperature after reaching 400°C for a few hours. This limitation could be partly due to the segregation process encountered by the substrate, leading to the apparition of lithium triniobate LiNb$_3$O$_8$ on its surface [7], but also to the degradation of the aluminium interdigital transducers (IDTs). Finding an electrode material for temperatures higher than 400°C is not trivial. Suitable candidates should gather the following properties:

- a high electrical conductivity
- a high Tamman temperature [8] which ensures the high-temperature physical stability of the electrodes
- a high resistance to oxidation
- a low density, allowing the generation of propagating SAW at high frequencies, i.e. in the 915 MHz or 2.45 GHz ISM bands

Unfortunately, none of the classical electrode metals can fulfil all the requirements. Aluminium has a very low electrical resistivity (2.6 µΩ.cm) and density (2.7 g/cm$^3$) but suffers from a low melting point (660°C) and poor resistance to oxidation, especially for long term monitoring. In the category of noble metals, platinum has an outstanding resistance to oxidation, a high melting temperature (1768°C), a reasonably low electrical resistivity (10.6 µΩ.cm), but a very large density (21.5 g/cm$^3$). Such a high value leads, when the frequency increases, to the increase of the reflection coefficient of the IDT fingers, and consequently hinders the SAW emission outside the IDT. This phenomenon may lead to the use of very thin Pt electrode fingers (around 50 nm at 2.45 GHz), with then several drawbacks: a large electrical resistance value and a high sensitivity to agglomeration phenomena leading to a very limited lifetime even at temperatures as low as 500°C [9].

Only an alloy could possibly gather the best of both worlds: the low electrical resistivity and density of Al enabling the generation and propagation of powerful high-frequency SAW signals, and the high melting temperature and resistance to oxidation of Pt, ensuring the electrodes to withstand harsh conditions. In the past, several alloys have been considered as thin film electrodes for high-temperature applications. Initially, they were based on noble metal alloys, Pt/Rh [2], or Ir/Rh [10] for instance. Those alloys offer very good thermal and chemical stability under air atmosphere at temperatures up to 800°C or even more. Unfortunately, they show at least two drawbacks. The first is, as for pure Pt, a very large density incompatible with R-DL applications. The second one is related to their high cost making them non compatible with industrial applications. In this context, Al-based alloys were considered as a good alternative from a physical and economical point of view. A first attempt was realized with RuAl thin film electrodes, showing very interesting performance up to 800°C in air atmosphere [11]. However, ruthenium is also a noble metal, and thus expensive. Moreover, the density of RuAl is too elevated, close to 8 g/cm$^3$, to be suitable for R-DL applications. More recently, TiAl alloy was investigated by the same group. Indeed, TiAl combines a high thermal stability with a very low density (4 g/cm$^3$), together with a lower cost than RuAl. However, Al and Ti have a high and rather close affinity for oxygen. Therefore, the oxide scale has a complex microstructure and fails at protecting the TiAl substrate above 400°C. This drawback can be circumvented by complexifying a bit the fabrication process, i.e. protecting TiAl films with a thin AlNO cover layer. In that case, TiAl IDTs show long term stability up to 500°C in air atmosphere [12].

In this context, another Al-based alloy, namely NiAl, could be very interesting for SAW R-DL applications. Indeed, NiAl alloy has a high melting temperature of 1640°C, a low density of 5.9 g/cm$^3$, a bulk electrical resistivity of 9 µΩ.cm and a very good resistance to oxidation in the bulk state [13-14]. Actually, in NiAl, aluminum oxidizes selectively with very slow parabolic growth kinetics ($k_p = 10^{-15}$ g$^2$.cm$^{-4}$.s$^{-1}$) [13]. Underneath the formed $Al_2O_3$ passivation layer, the NiAl phase remains stable since its stability domain is large (around 10 at. %) [15]. However, the high-temperature behavior of very thin NiAl films, with a thickness in the hundreds of nanometers range, is mostly unknown.

Consequently, this study has two main goals. Firstly, to re-examine thoroughly the high-temperature limitations of the $LiNbO_3$-based SAW R-DL sensors technology, and in particular to determine if the congruent $LiNbO_3$ segregation process is prohibitive for applications above 350°C. Secondly, to examine the potential of NiAl IDTs for long-term operations at temperatures up to 500°C.

## II. METHODS

### A. Fabrication

The whole study was conducted on 128° rotated Y-cut $LiNbO_3$ substrates (Neyco, Vanves, France) whose large electromechanical coupling coefficient ($K^2$) of 5.5% in the X direction (for the Rayleigh wave) is particularly suitable for the R-DL technology.

150 nm-thick metal or alloy films were sputtered (Alliance concept DP650, Annecy, France) on the substrates: Al and Pt as references, and NiAl as a promising high-temperature electrode candidate. Sputtering parameters for NiAl films are summarized in Table I.

Connected reflective delay lines (Fig.1) [16] were then realized by photolithography, followed by chemical etching for Al and NiAl electrodes, and ion beam etching in the case of Pt. The connected R-DLs were made of three IDT benches, each composed of 11 pairs of fingers, with a 50% metallization ratio.

TABLE I. SPUTTERING PARAMETERS FOR NICKEL-ALUMINIUM ALLOY FILMS

| TARGET | NiAl (50/50 at. %) ; pur.: 99.99% |
|---|---|
| SUBSTRATE HEATING | No intentional heating |
| SPUTTERING POWER | 100 W RF |
| PLASMA PRESSURE | 3.10$^{-3}$ mbar |
| SPUTTERING GAS | Ar (16 sccm) |

The wavelength was 9.2 µm, leading to an operating frequency close to 433 MHz. This frequency was chosen in order to make a comparison between the respective behaviors of Al, Pt and NiAl-based R-DLs. Indeed, as explained previously, frequencies in the GHz range would require Pt films with a thickness under 100 nm, which would be detrimental to their performance.

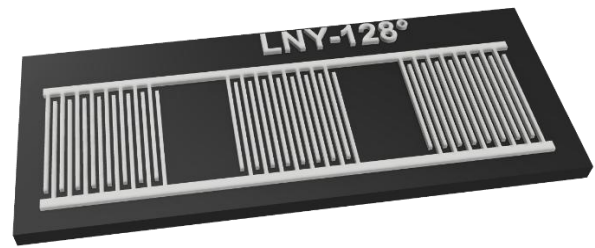

Fig. 1. Schematic view of reflective delay lines (RDL) with 3 connected IDTs design [16].

### B. Characterisations

First, all the different samples were annealed in air atmosphere for 48h at 500°C. They were electrically characterized before and after this annealing treatment using a RF probe station (PM5 SÜSS MicroTec SE, Garching, Germany) and a vector network analyser (VNA, Agilent-N5230A). The changes in the microstructure of the electrodes and the substrate were determined by X-ray diffraction method (XRD) in Bragg–Brentano geometry (Brucker D8 Advance — CuK$\alpha$1 : $\lambda$ = 1.54056 Å).

Then one device with NiAl IDTs was tested for a much longer time, namely 250h at 500°C, again in air atmosphere. This device was continuously electrically characterized in-situ the entire time using a dedicated setup described below.

Finally, at the end of this 250h-long period, the structure of the device was examined at the nanoscale level by transmission electron microscopy (TEM) and scanning transmission electron microscopy (STEM), using a JEM- ARM 200F Cold FEG TEM/STEM (JEOL Ltd., Tokyo, Japan), operating at 200 kV and equipped with a spherical aberration (Cs) probe corrector (point resolution 0.19 nm in TEM mode and 0.078 nm in STEM mode). Chemical compositions were determined using energy dispersive X- Ray spectroscopy (EDXS). EDXS spectra were recorded in STEM mode by means of a Dry SD Jeol spectrometer (SDD). TEM lamellas were prepared by focused ion beam (FIB) method, using a FEI Helios Nanolab 600i (FEI, Hillsboro, Oregon, USA).

### C. In-situ electrical measurements

A homemade in-situ electrical measurement setup (Fig. 2) was designed in order to monitor the SAW devices response continuously at high temperature during several hundreds of hours. Stainless steel tips were homemade in order to make contact with the device under investigation. To do so, the tips and the underlying device are clamped onto a ceramic plate. Both tips are connected respectively to the stainless-steel shielding and inner conductor of a high-temperature RF cable (Meggitt, Irvine, CA) via screw terminals. The RF cable is connected to a VNA which is controlled automatically by a computer via a GPIB port. The entire system makes it therefore possible to automatically record the device impedance at regular intervals of time.

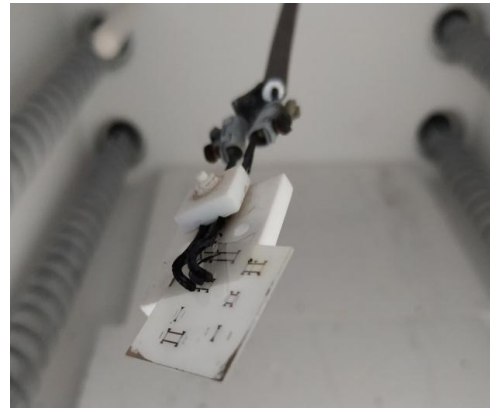

Fig. 2. Photograph of the setup used for high-temperature in-situ electrical measurements.

## III. RESULTS AND DISCUSSIONS

### A. Short-term annealing results

Figure 3 shows the evolution, after a 48h-long annealing at 500°C in air atmosphere, of the time-resolved $S_{11}$ reflection coefficient of three SAW R-DLs using platinum, aluminium and NiAl electrodes respectively.

In the case of aluminium, we can observe that the devices response is strongly deteriorated after the annealing, the pulse amplitudes decreasing by at least -7dB. This degradation was expected and is likely related, as already mentioned hereinabove, to the partial oxidation of the Al IDTs, but also to the potential segregation process encountered by the LiNbO$_3$ substrate. On the contrary, the pulse amplitudes of the Pt-based devices are improved by +1dB at the end of the annealing process. This amelioration is attributed to the recrystallization of the platinum thin film [17]. This is a very important result as it shows that, at least at 433 MHz, the LiNbO$_3$ segregation process does not seem to have a significant adverse effect on the sensor response. The TEM investigations described further help to understand this result.

The characterization of the NiAl-based devices confirms the high potential of this alloy for the targeted applications. Indeed, as for Pt IDTs, the pulse amplitudes are slightly improved averagely by +3 dB during the 48h-long annealing process at 500°C. XRD characterizations show a recrystallization of the NiAl film that could explain this phenomenon (Fig. 4). However, this analysis also evidences the apparition of the $\alpha$-Al$_2$O$_3$ phase, indicating likely the partial oxidation of the IDTs. It seems that, at least for an operating frequency of 433 MHz, and for an annealing duration of 48h, this deterioration process of the NiAl electrodes is not detrimental to the device performances, or is counterbalanced by the recrystallisation process. In-situ long-term characterizations, described in the next paragraph, will allow to determine if this oxidation process evolves, deteriorating more and more the device performance, or if, on the contrary, a passivation layer forms and protects the electrode from further oxidation. Finally, we can also observe on the XRD spectrum the presence of lithium triniobate LiNb$_3$O$_8$. There again, longer annealing processes will reveal the impact of the associated segregation process on the device lifetime.

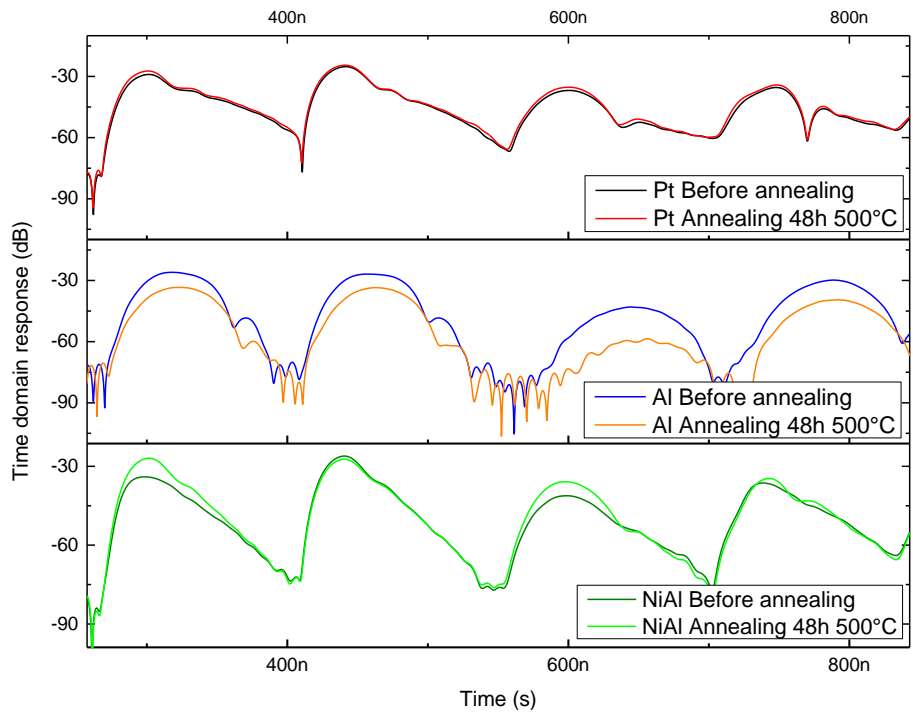

Fig. 3. Reflection coefficient ($S_{11}$) in the time domain before and after 48h at 500°C for SAW R-DLs using Pt, Al and NiAl as IDTs material, respectively.

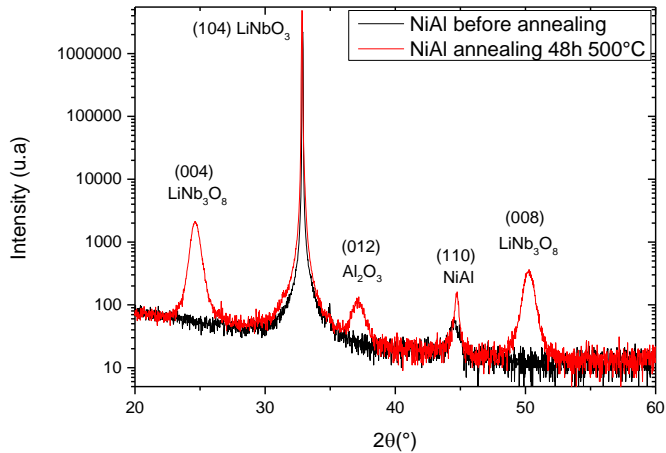

Fig. 4. XRD diagram of NiAl/LiNbO$_3$ R-DL before and after 48h at 500°C.

### B. Long-term annealing results

The former results show that the SAW signal of the devices based on NiAl IDTs is intact (and even slightly improved) after 48h at 500°C. In order to determine the actual potential of NiAl/LiNbO$_3$ R-DLs, a device was continuously in-situ electrically characterized for 250h at 500°C, using the dedicated setup described hereinabove. To track the variation of the frequency versus temperature, a gate function was applied in the time domain (Fig. 3) to consider only one peak. After applying the fast Fourier transform (FFT), phase versus frequency is plotted and the frequency corresponding to a constant phase is tracked at different temperatures and plotted versus time. Figure 5 shows simultaneously the temperature profile and the evolution of the device operating frequency throughout the experiment. It was observed during the initial heating phase that the resonance frequency evolves very linearly with the temperature. A temperature coefficient of frequency (TCF) of -78.6 ppm/°C was calculated, in very good agreement with the value usually reported in the literature for the Y128°-X direction of LiNbO$_3$. Then, during the plateau at 500°C for 250 hours, the resonance frequency remains very stable, the calculated standard deviation being 450 Hz for a resonance frequency close to 433 MHz, corresponding to 1.04 ppm. Thus, if the device was used as a wired temperature sensor, the temperature standard deviation would be of 0.01°C.

The observation of the sensor response in the time domain, measured at different moments of the experiment, gives even more information on the sensor capabilities (Fig. 6).

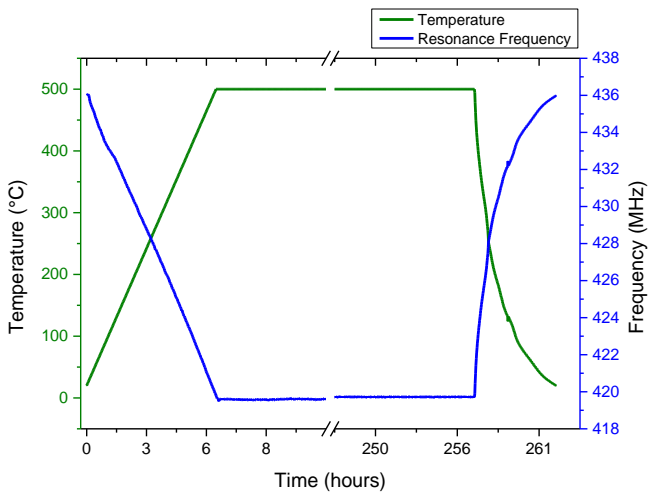

Fig. 5. Resonance frequency tracking (blue curve) of NiAl/LiNbO$_3$ R-DL during long-term high-temperature experiment. The green curve describes the temperature profile during the experiment.

Firstly, it appears that the time pulses amplitude at 500°C, at the start of the plateau, is very close to that at room temperature, at the beginning of the experiment. However, several phenomena should make high-temperature signal weaker than the one at room temperature. One can cite for instance the increase of the substrate conductivity with the temperature [18], the augmentation of the IDTs resistivity because of their metallic nature, leading to stronger losses by Joule effect, or else the thermal phonon density rising with the temperature, causing more scattering of the phonon induced by the IDTs. It is likely that all those phenomena are counterbalanced by the IDTs recrystallisation that likely happens during the initial heating phase during which the temperature is increased between the ambient and 500°C.

A second important result appears in the upper part of figure 6. At the end of the experiment, after this 10-days period at 500°C, the SAW signal at room temperature is quite identic to that at the very beginning of the experiment. It is similar regarding the position of the time pulses, confirming that the sensor capabilities are entirely saved. Actually, the main change concerns the pulses amplitude, which is even slightly better at the end of the experiment. This result definitively confirms the capability of NiAl/LiNbO$_3$ R-DLs to withstand high temperatures up to 500°C, for very long durations.

However, some aging of the device can be seen through the lower part of figure 6, which depicts the SAW signal measured at different times during the plateau at 500°C. One can notice a slight degradation of the amplitude of the time pulses by 3 dB during the first 50 hours. Then, the aging slows down sharply as additional losses of only 1 dB can be observed between the 50$^{th}$ and the 250$^{th}$ hour of the annealing at 500°C. This means actually that the sensor structure and performance are practically stabilized after 50 hours of exposure at 500°C. It is likely that the kinetics of the IDTs oxidation and the substrate segregation evidenced by XRD (Fig. 4) become very slow at the end of the first two days of annealing at 500°C. Besides lithium niobate segregation, the slight signal degradation can be linked to the metal thickness diminution, which can cause a decrease of K$^2$ and an increase of IDT's resistance due to skin depth effects.

Cross-section TEM images enable a better understanding of this aging process (Fig. 7). The IDTs structure appears deeply modified after the 250h-long annealing process at 500°C. Indeed, the initial 150 nm-thick NiAl film is transformed into a three-layers stacking. STEM-EDXS analyses show that the overlayer (22 nm-thick), as well as the bottom layer (15 nm-thick), contains only Al and O atoms (Fig. 8). Both are identified as $\alpha$-Al$_2$O$_3$ thin films using selected area electron diffraction (SAED) patterns (not shown here). In between, a 95 nm-thick NiAl layer remains. Based on the above presented in-situ electrical characterisations, it is likely that the Al$_2$O$_3$ over- and bottom layers mainly form during the first 50h of annealing at 500°C, acting as passivation layers that protect the inner NiAl layer from further oxidation.

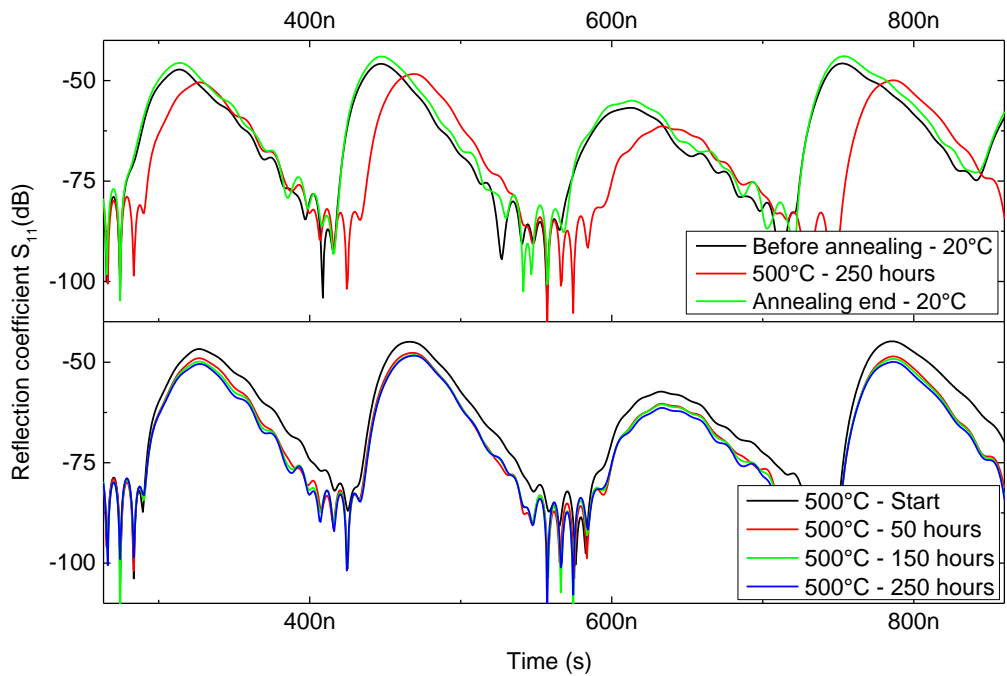

Fig. 6. Reflection coefficient ($S_{11}$) in the time domain of a NiAl/LiNbO$_3$ R-DL a) before, after and during a 250 hours annealing period at 500°C, b) at various moments at 500°C

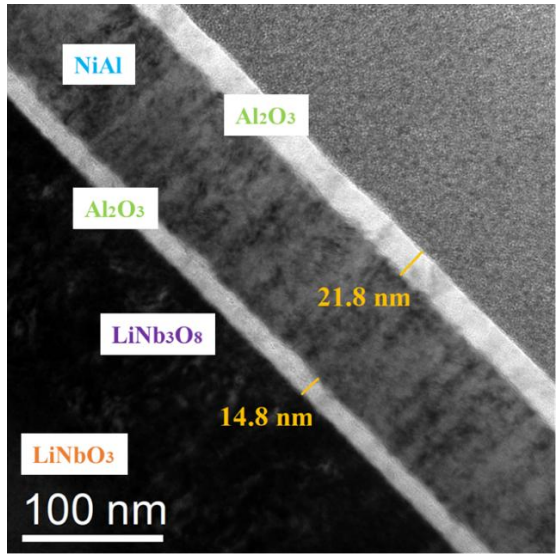

Fig. 7. TEM Bright Field micrograph of NiAl alloy thin film on LiNbO$_3$ substrate after 250 h at 500°C

STEM images bring additional information, showing in particular that an alteration of the surface of the congruent LiNbO$_3$ occurs during the annealing process. This phenomenon is located in the first 150-200 nm of the substrate (Fig. 8, picture in the top left-hand corner). SAED patterns reveal that this overlayer is constituted by the LiNb$_3$O$_8$ phase (Fig. 8). EDXS-mapping done at the interface at higher magnification supposes that this overlayer is not homogeneous with the apparition of pores during the segregation process (Fig 9). The thickness of this deteriorated area at the surface of the substrate is much smaller than the wavelength of the devices (9.2 μm), and thus than the penetration depth of the wave, which probably explains that the segregation process has no significant influence on the device performance (Fig. 6). It will be necessary to check if it remains true with devices operating in the 915 MHz or 2.45 GHz ISM bands, whose wavelengths are reduced to 4.2 μm and 1.6 μm respectively. If not, stoichiometric LiNbO$_3$ substrates could constitute an ideal alternative, as these crystals are expected to be thermodynamically stable up to their melting temperature at 1170 °C [19].

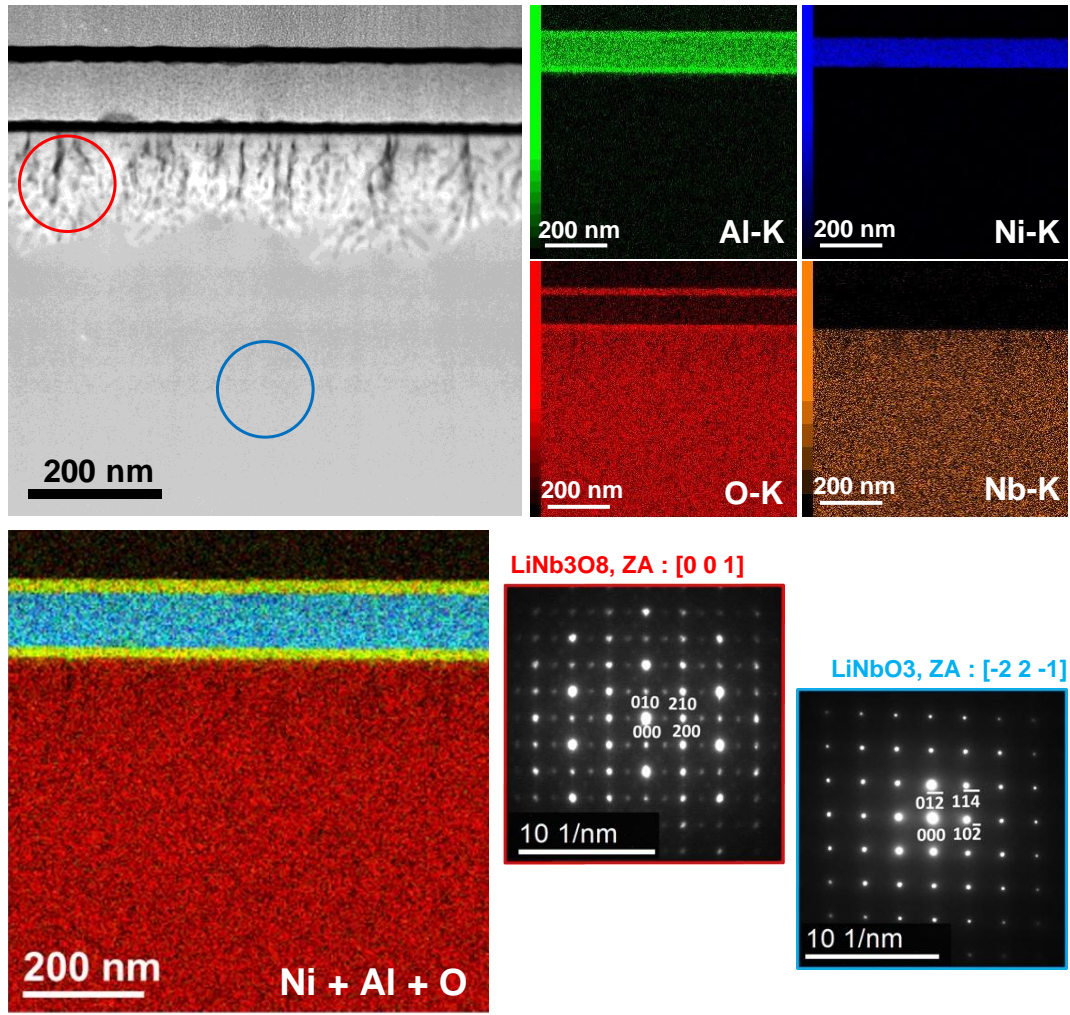

Fig. 8. STEM HAADF (High Angle Annular Dark Field ) micrograph and corresponding EDXS-maps of oxygen (red), aluminum (green), nickel (blue), niobium (orange) of NiAl devices after a 250 hours annealing at 500°C, and SAED patterns of the deteriorated area (red circle) and LiNbO$_3$ substrate (blue circle)

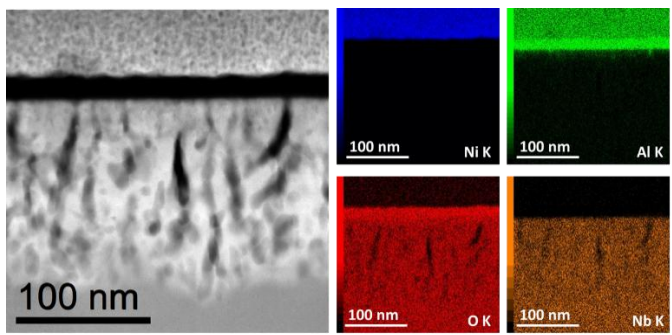

Fig. 9. STEM HAADF micrograph with EDXS-maps of nickel (blue), aluminium (green), oxygen (red) and niobium (orange) of NiAl devices after a 250 hours annealing at 500°C

## IV. CONCLUSIONS AND OUTLOOKS

In this study, an original alloy, namely NiAl, was tested as a promising material to make IDTs for high-temperature, long-term R-DL applications up to 500°C. Indeed, NiAl combines in the bulk state all the requested properties for the targeted applications: high electric conductivity, low density, high melting temperature and great resistance to oxidation. This study thus aimed at investigating the behavior of NiAl thin films at high temperatures up to 500°C, regarding SAW R-DL applications. Another goal was to determine if the segregation process undergone by congruent LiNbO$_3$ substrates at high temperature is prohibitive for the targeted applications.

48h-long annealing process conducted at 500°C under air atmosphere confirmed the high potential of NiAl thin films. Indeed, the time-resolved S$_{11}$ response of R-DLs operating at 433 MHz was slightly improved in the case of NiAl or Pt IDTs, whereas it was strongly deteriorated for Al IDTs. However, XRD analyses revealed the partial oxidation of NiAl films, together with the apparition of lithium triniobate $LiNb_3O_8$. The capabilities of NiAl/$LiNbO_3$ R-DLs in the long term were then investigated during a 250h-long annealing at 500°C. The results are particularly promising. Indeed, a frequency standard deviation of 1.04 ppm only was observed during the long plateau at 500°C. Moreover, at the end of the experiment, after more than ten days at 500°C, the position of the time pulses at room temperature was the same as that at the very beginning of the experiment, while their amplitude was a bit improved. Continuous in-situ electrical monitoring of the device throughout the experiment showed that some aging happened, mainly during the first 50h at 500°C. TEM investigations revealed that the great resistance of NiAl films to high temperatures up to 500°C is related to self-passivation: 22 nm-thick and 15 nm-thick $Al_2O_3$ films respectively formed above and below a remaining 95 nm-thick NiAl layer. TEM analyses also revealed that the segregation process is located in the first 150-200 nm of the substrate. This thickness is much smaller than the wavelength (9.2 µm) used in this study, which explains that this phenomenon had no significant impact on the device performance.

Additional experiments are currently ongoing to explore the behavior of NiAl/$LiNbO_3$ R-DLs (i) at temperatures higher than 500°C (ii) in the 915 MHz and 2.45 GHz ISM bands, much more adapted than the 433 MHz band to the R-DL technology. It is possible that, as the ratio between the deteriorated thickness of the substrate and the wavelength increases, the segregation process becomes an issue. In that case, stoichiometric $LiNbO_3$ substrates should constitute an alternative, as these crystals are thermodynamically stable up to their melting temperature.

### ACKNOWLEDGMENT

This work was carried out within the framework of the French PIA "Lorraine University of Excellence" project (ANR-15-IDEX-04-LUE). Authors would like to thank Sylvie Migot for the preparation of FIB lamellas for TEM characterizations and the Center of Microscopy, Microprobe and Metallography (CC3M) of Institute Jean Lamour (IJL) for FIB and TEM facilities.

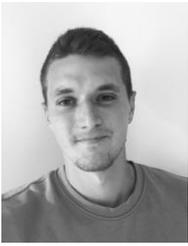

**J. Maufay** received a Msc degree in the field of Engineering and Optics from CentraleSupélec, France in 2019. Then he started a PhD degree in Nano- and Microtechnologies at the LMOPS (Laboratoire Matériaux Optiques, Photonique et Systèmes) and Institut Jean Lamour of Universite de Lorraine. His PhD project is focused on the development of high-temperature SAW sensors based on lithium niobate crystals.

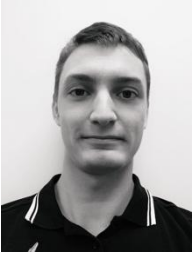

**B. Paulmier** received an Engineering degree from ESIREM, Dijon, France in 2019. He is currently working towards his PhD Degree from Université de Lorraine at Institut Jean Lamour, developing surface acoustic wave sensors for Structural Health Monitoring industrial applications.

**M. Emo** received an engineering degree in the field of Analytical Sciences from University of Strabourg, France in 2009. Since 2010, she is working as an engineer at the French National Centre for Scientific Research (CNRS) and since 2019 as a microscopist at Institut Jean Lamour of Université de Lorraine.

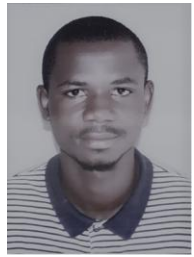

**U. Youbi** obtained a Msc degree in photonics and optics for materials conjointly from Université de Lorraine and CentraleSupélec, France in 2021. Since October 2021, he has been a Ph.D. student in the Micro and nanosystems team at Institut Jean Lamour. He studies wireless SAW sensors for high temperature applications.

**E. Mbina** received a Master degree in the field of photonics and optics for materials from CentraleSupélec and Université de Lorraine, France in 2022. He worked in 2022 as an intern in the micro- and nanosystems team of the Institut Jean Lamour, on the development of SAW sensors for high temperature, based on NiAl electrodes and lithium niobate substrate.

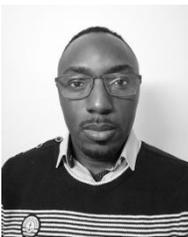

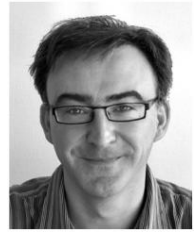

**T. Aubert** received the M.Sc. degree in Physics and the Ph.D. degree from the University of Nancy I, France, in 2007, and 2010 respectively. He then has been an Associate Professor with the Université de Savoie (2011-2014), Annecy, France, and with CentraleSupélec (2014-2020), Metz, France. Since 2020, he is a full professor at Université de Lorraine, Metz, France. He is currently the director of the LMOPS (Laboratory of Optical Materials, Photonics and Systems). His research activities are focused on the investigation of new materials and structures for high-temperature SAW applications (WLAW structures, Ir/Rh thin film electrodes, piezoelectric ScAlN thin films, congruent and stoichiometric lithium niobate crystals, NiAl thin film electrodes).

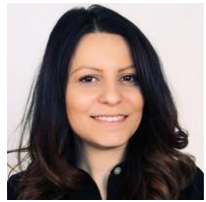

**N. Kokanyan** is associate professor at the LMOPS laboratory, CentraleSupélec, Metz campus. She received her PhD in physics from the University of Lorraine in 2015, specializing in polarized Raman spectroscopy of photorefractive Lithium Niobate crystals. Her current research focuses on Raman spectroscopy applications in optics, biotechnology, and structural materials. In addition to her research, she serves as the Academic Dean on the Metz campus, overseeing the engineering curriculum and as in charge of Innovation & Intrapreneurship branch at CentraleSupélec Metz campus. Ninel Kokanyan is also the referent of gender equality on campus

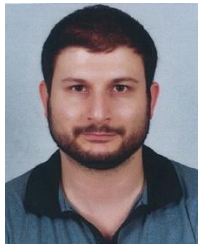

**S. Hage-Ali** (M'10) was born in Strasbourg, France, in 1982. He received an Engineering degree from Ecole Centrale de Lille, Villeneuve-d'Ascq, France, the M.S. degree in micro-nanotechnology and the master's degree in international projects engineering from the Université de Lille 1, Villeneuve-d'Ascq, in 2005 and 2006, respectively, and the Ph.D. degree in micro-nanotechnology, acoustics and telecommunications from Ecole Centrale de Lille, in 2011.

He was a Postdoctoral Fellow with the University of Illinois at Urbana–Champaign, Champaign, IL, USA. Since 2014, he has been an Associate Professor with Université de Lorraine, Nancy, France, and the Micro-Nanosystems Group, Institut Jean Lamour, Nancy. His research interests include flexible/stretchable electronics, micro-nanosystems, microwaves, antennas, and surface acoustic wave sensors.

Dr. Hage-Ali received a Fulbright Grant in 2011. He is a founder and past president of the IEEE France Section Sensors Council Chapter and its current treasurer.

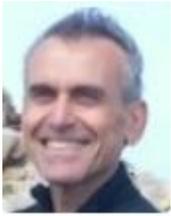

**M. Vilasi** is a Full Professor at Université de Lorraine, Nancy, France within Jean Lamour Institute, specialized in Inorganic Chemistry and Material Science. His activities are focused on new intermetallics for high temperature corrosion. 1986 – 1989: development of a NiCoCrAlYTa coating resistant to the oxidation in air (1000°C) and to the corrosion in molten $Na_2SO_4$ (850°C)//1989 – 1998: study of b-Ni(Pd)Al coating for turbine blade application; development of new niobium alloys (Nb-Ti-Al) and new silicide coatings (Nb-Fe-Cr-Si) for niobium alloys// since 1998: development of metallic and ceramic materials for tools used in i) SMR atmospheres, ii) in molten glass (Ni-base superalloys, $MoSi_2$-base materials, $SnO_2$ and $ZrO_2$- base materials) and iii) in Na and Pb-Bi liquid metal (T91, Inox 316).

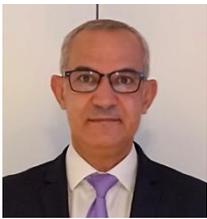

**Omar ELMAZRIA** is a Full Professor at Université de Lorraine, Nancy, France within Jean Lamour Institute and he was guest Professor at several Universities around the world (SFU, Canada; IoA, Chinees Academy of Sciences; UCF, USA, SJTU, Shanghai). O. Elmazria is the head of Nanomaterial, Electronic and Living department within the IJL and his current research focuses on SAW devices for sensing applications. He is member of Technical Program Committee of several international conferences and society including IEEE IUS, IEEE MTT-26-RFID-Wireless-Sensor-and-IoT; SAW Symposium; IFTC. He is also AdCom member of IEEE UFFC Society and IEEE RFID Council. In 2017, he was a recipient of the URSI-France medal from the International Union of Radio Science.